\documentclass{jcgt}

\usepackage{amsmath,amsfonts,amssymb}
\usepackage{listings}
\usepackage{booktabs}
\usepackage[hangindent=12pt]{subfig}
\makeatletter
\def\do@url@hyp{\do\-\do\_}
\makeatother

\setciteauthor{Bartlomiej Wronski}
\setcitetitle{GPU-Friendly Laplacian Texture Blending}

\accepted{2023-03-28}{2025-01-09}{2025-02-19}{Angelo Pesce}{14}{1}{21}{39}{2025}
\seturl{http://jcgt.org/published/0014/01/02/}

\begin{document}

\title{GPU-Friendly Laplacian Texture Blending}

\author
       {Bartlomiej Wronski\\NVIDIA, USA
       }

\teaser{
  \includegraphics[width=\linewidth]{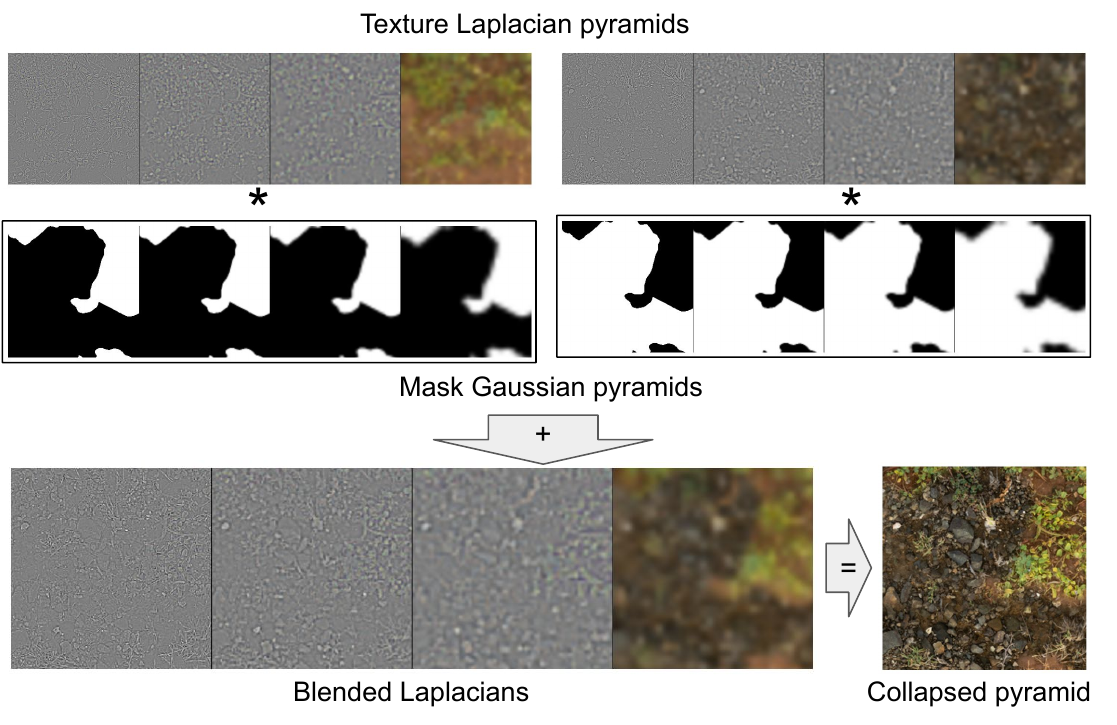}
  \caption{Overview: Instead of blending material textures with a fixed blending radius, we propose to blend different Laplacian pyramid levels with different mask sharpness proportional to Laplacian feature size.
  This ensures contrast and detail preservation as well as smooth perceptual transition.
  Laplacian levels are constructed in place from traditional texture mipmaps.}
  \label{overview}
}

\maketitle
\thispagestyle{firstpagestyle}

\begin{abstract}
\small
Texture and material blending is one of the leading methods for adding variety to rendered virtual worlds, creating composite materials, and generating procedural content.
When done naively, it can introduce either visible seams or contrast loss, leading to an unnatural look not representative of blended textures.
Earlier work proposed addressing this problem through careful manual parameter tuning, lengthy per-texture statistics precomputation, look-up tables, or training deep neural networks.
In this work, we propose an alternative approach based on insights from image processing and Laplacian pyramid blending.
Our approach does not require any precomputation or increased memory usage (other than the presence of a regular, non-Laplacian, texture mipmap chain), does not produce ghosting, preserves sharp local features, and can run in real time on the GPU at the cost of a few additional lower mipmap texture taps.
\end{abstract}

\section{Introduction} 

Texture and material creation is one of the most time-consuming aspects of 3D content creation and defines the final appearance of the rendered objects.
In physically based shading, the artist defines the surface reflectance and other physical properties of the BRDF~\cite{burley2012physically}.
To help artists author physically based materials, a common industry practice is creating a hierarchy of progressively more complex materials that get blended through masking~\cite{neubelt2013crafting}.
Masking is typically manually tweaked by artists but can be extended to the procedural generation of infinite materials.
To reduce visible repetitiveness and material tiling, \citet{heitz2018high} proposed hexagonal macro-tiling of random rotations of material textures.
They analyzed a common problem of naively blending textures---either sharp, unnatural transitions or contrast loss and ghosting---and proposed a solution based on local histogram correction based on precomputation.
We propose a different solution to the problem of texture and material blending that can also be applied to interpolate different textures.
Our solution blends local Laplacians to reduce variance loss and provide a perceptually natural transition of differently sized features.
Laplacian pyramid construction is approximated inline in the final shader using traditional texture mipmaps (used for regular trilinear filtering) and requires no costly precomputation.
Our approach is designed to run in real time on the GPU and requires defining only a single parameter: the number of the Laplacian levels that affect the maximum size of the features that get blended.

\section{Related Work} 

Texture blending is common in practice and is part of texture and material creation, but relatively little of it has been explored by the literature.
Artists are assumed to tweak blending masks and source materials until the desired look is achieved.
The state-of-the-art procedural texturing tool Adobe Substance Designer uses various simple pointwise blending modes to facilitate that process~\cite{substanceblending}.
\citet{Mikkelsen2022Hex} proposed to use simple pointwise blending with a manually designed, content-dependent weight curve for natural-looking transitions.

While artists can manually adjust masks and the appearance of materials in traditional workflows, procedural texture synthesis aims to automate this process.
Pure pointwise operations often fail to produce reliable and consistent procedural blending results, as a single pixel does not inform about neighbors, image patterns, or structures.
Most procedural texture synthesis publications rely on global or local neighborhood approaches during runtime or as a precomputation step.

One of the earliest practical works was noise by example~\cite{galerne2012gabor}.
Those early methods were costly and required offline optimization procedures, such as basis pursuit.
Subsequent works targeted performance optimizations and reduction of the precomputation needed.
\citet{yu2010lagrangian} proposed global variance-based normalization for blending fluid textures. 
\citet{heitz2018high} identified this approach's quality shortcomings and instead proposed adjusting the local texture values based on offline precomputation of optimal transport of histograms.
In later work, \citet{burley2019histogram} proposed simplifying that process through 1D precomputation along with improvements to reduce visual artifacts by taking clipping into account.
Recently, \citet{fourniersauvage2024mixmax} introduced a novel pointwise operator combined with fast precomputation techniques that guarantee consistent minification, antialiasing upon magnification, and stationarity of the resulting blended textures.

\enlargethispage{3pt}

\section{Laplacian Pyramid Blending}

Blending natural and photographic images is a common operation in image processing literature, focusing significantly on perceptual effects on image color, details, and discontinuities. \citet{perez2003poisson} have approached the problem of blending images in the gradient domain by swapping image gradients and solving a screened Poisson equation. One of the key insights of their method is that local image gradient discontinuities don't lead to a perceived image discontinuity.

In parallel and in a similar manner, \citet{brown2003recognising} proposed to blend panorama photographs using a simple, two-level frequency decomposition of the image---a \textit{detail} layer with high frequencies that are blended locally, and a \textit{global} layer that is blended with a large radius to prevent visible seams or discontinuity.
\citet{efros2005imagepyramids} expanded this idea to a multi-level Laplacian decomposition and the blending of images using Laplacian pyramids in his influential course on image processing.

This technique is commonly used in perceptual image processing: for example, in the Exposure Fusion~\cite{mertens2007exposure} algorithm used for high-dynamic-range (HDR) fusion of multiple low-dynamic-range images or to locally tone-map an HDR image in the HDR+ pipeline~\cite{hasinoff2016burst}.
Earlier HDR fusion approaches would, for example, apply the tone mapping only to the bilaterally filtered image while adding the non-tone-mapped \textit{detail} layer back later, which often resulted in an unnatural, exaggerated, and over-detailed look.
Exposure Fusion solved this problem with a much smoother, layer-dependent blending radius from Gaussian-blurring blending masks.
We analyze why this method preserves contrast from the image statistics and signal-processing perspectives and propose blending materials and their textures using Laplacian pyramids.

\begin{figure}[t]
\includegraphics[width=1.0\linewidth]{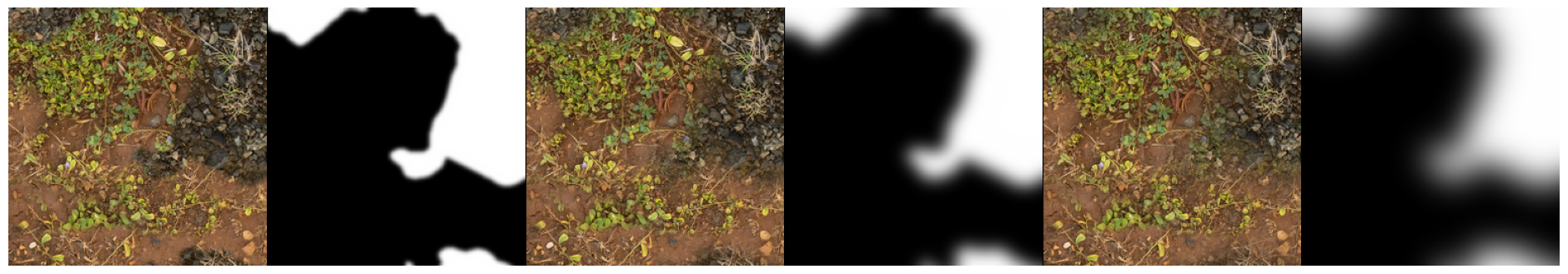}
\caption{Different blending radii can affect the visual look of blended tiled textures. We can observe either unnatural, harsh transitions (left) or significant blurriness and contrast loss (middle). Furthermore, the rightmost example shows significant ghosting and overlap of distinct details.}
\label{fig_intro_blend}
\end{figure}

\section{Method}
When directly linearly blending different textures, the transition radius and mask smoothness affect the final image look (Figure~\ref{fig_intro_blend}).
The rightmost example shows the smoothest transition but has less contrast and detail than the two others, with smaller blending radii.
Furthermore, various small visual features overlap, producing an unnatural \textit{ghosting} effect.
Addressing the shortcomings of such blending was one of the contributions of the method of \citet{heitz2018high}, which was one of the inspirations for our work.

We propose a different and straightforward method, based on the ideas present in the image processing work: blending Laplacians with different radii~\cite{brown2003recognising,efros2005imagepyramids,mertens2007exposure}.
We take images $x$ and $y$ and decompose them with Laplacian operators:
\begin{align}
x &= l_{x0} + L_{x1} + L_{x2} + \cdots + G_{xn},\\
y &= L_{y0} + L_{y1} + L_{y2} + \cdots + G_{yn},
\end{align}
where $L_{xm}$ is a Laplacian pyramid level of the image $x$ and $G_{xn}$ is the final Gaussian level of the signal.
Similarly, we create multiple \textit{Gaussian} levels of the mask image $m$: $G_{m0}, \ldots, G_{mn}$.
The radii of Gaussian blurring of the consequent Gaussian levels are assumed (but not required) to be the same as the radii of Gaussian blurs during the construction of Laplacian pyramids for images $x$ and $y$.

\begin{figure}[tb]
\centering
\includegraphics[width=0.8\linewidth]{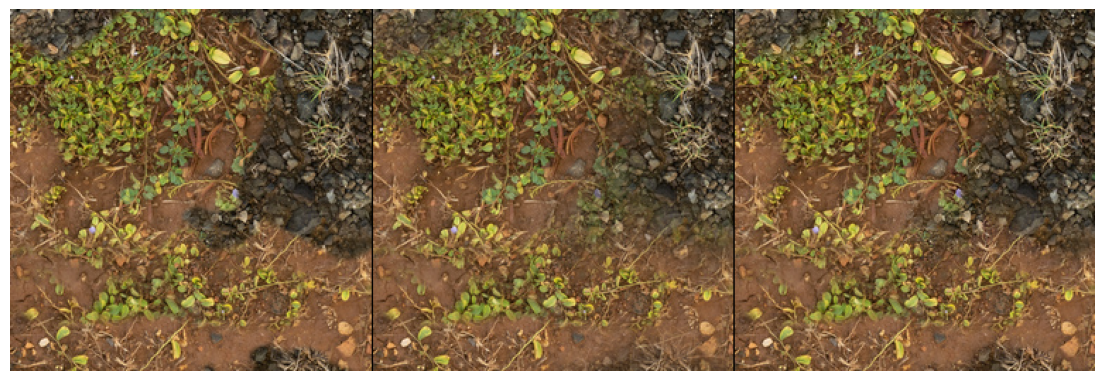}
\caption{Left: Direct texture blending with a small radius. Middle: Direct texture blending with a large radius. Right: Laplacian texture blending using three Laplacian levels with a wide radius blending low-frequency details and a narrow radius blending high-frequency details.}
\label{intro_laplacian_blending}
\end{figure}

Given this notation, the proposed blending operation is as follows:
\begin{equation}
\mathit{Blend}(x,y,m) = G_{xn}\cdot G_{mn} + G_{yn}\cdot (1-G_{mn}) + \sum_{i=0}^{n-1} L_{xi}\cdot G_{mi} + L_{yi}\cdot (1-G_{mi}).
\end{equation}
A visual example of our method's appearance is presented in Figure~\ref{intro_laplacian_blending}.

This method preserves sharpness and contrast in the transition area while blending the textures over a large area.
This operation generalizes to blending more than just two textures. In such cases, we replace the $m$ and $(1-m)$ with different masks $m_0, \ldots, m_k$ and add the weighted Laplacians linearly.
The number of added Laplacian levels defines the sharpness of the transition.
Before showing the impact of the Laplacian count, the pyramid construction filters, and presenting a practical implementation, we analyze why the method works well for preserving the visual appearance and contrast of the blended textures.

\subsection{Perceptual Impact of Frequency-Dependent Blending Radius}
\begin{figure}[b]
\centering
\includegraphics[width=\linewidth]{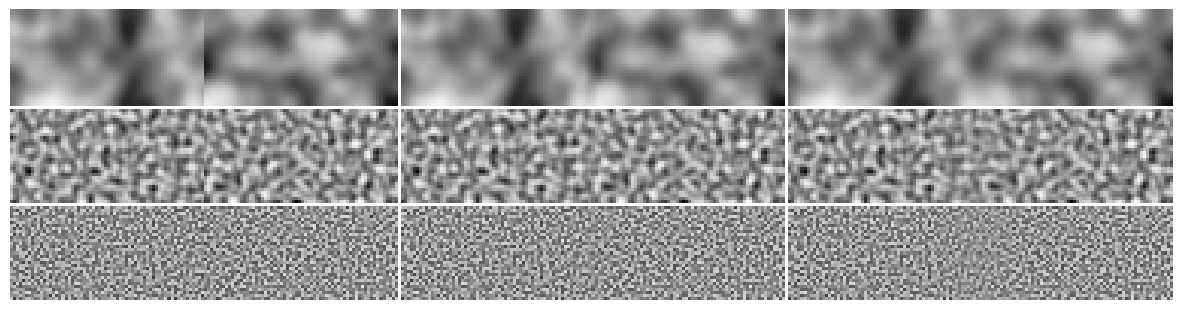}
\caption{Top to bottom: Noise textures with a different frequency content, from low to high frequencies. Left to right: Different blending radii between two textures. Different frequencies of noise require different transition radii for the most natural appearance. A small blending radius produces visible discontinuities on low-frequency content, while a wide radius causes contrast and detail loss on high-frequency textures.}
\label{noise_octaves}
\end{figure}
\citet{efros2005imagepyramids} noted in his course how differently sized features require different blending radii to \textit{look natural}.
We demonstrate this effect on noise textures in Figure~\ref{noise_octaves}.
When the transition radius is mismatched with the frequency content of the noise texture, it results in a discontinuous look or a visible blurry \textit{stripe} between the two textures.
A perceptually optimal blending radius is wide for low-frequency noise and medium for medium-frequency noise, and there is almost no transition for high-frequency noise.

The reason for visible discontinuities when using a small blending radius to blend low-frequency features can be analyzed from a signal-processing perspective.
Multiplying a texture by a mask in pixel space is the same as the convolution of two signals in the frequency space.
A harsh transition ramp has rich frequency content and, after multiplication, causes new, high frequencies not present in either of the original images to appear.

It would seem that a wide radius---with a smooth falloff---would be preferred, but it's not the case for medium and high frequencies.
We look at those cases from the perspective of variance loss.

\subsection{Analysis: Variance Loss}
We use the variance of the color variation in the texture as a simple-to-analyze proxy for the contrast.
We look at the texels of blended textures $x$ and $y$ as random variables $X$ and $Y$ and analyze their blended variance $\mathit{Var}(a\cdot X + (1-a)\cdot Y)$ after linear blending:
\begin{equation}
\mathit{Var}(a\cdot X + (1-a)\cdot Y) = a^2\cdot \mathit{Var}(X) + (1-a)^2 \cdot \mathit{Var}(X) + a \cdot(1-a) \cdot \mathit{Cov}(X, Y).
\end{equation}
The effect on variance is zero on the edges of the transition and the largest in the middle of the transition:
\begin{equation}
\mathit{Var}\left(\frac{X}{2} + \frac{Y}{2}\right) = \frac{\mathit{Var}(X)}{4} + \frac{\mathit{Var}(X)}{4} + \frac{\mathit{Cov}(X, Y)}{2}.
\end{equation}
In the case of uncorrelated blended textures, blending reduces the variance by half, reducing visual contrast.
If the textures are anti-correlated (which can happen when blending the same periodic texture with different phase offsets), it can lead to the complete zeroing of the texture variation and detail.

The blending radius defines the size of the area with a lowered variance, which for uncorrelated variables with the same variance on average is equal to
\begin{equation}
\int_0^1 a^2\cdot \mathit{Var}(X) + (1-a)^2 \cdot \mathit{Var}(X) \mathrm{d}a = \frac{2}{3}\mathit{Var}(X).
\end{equation}
Outside of the blending area, there is no variance loss.
The wider the transition area, the more variance and contrast are reduced.
While looking at variance only and restoring it is insufficient~\cite{heitz2018high}, and better methods operate on full histograms, it leads naturally to the analysis of benefits of the Laplacian decomposition.

\subsection{Laplacian Decomposition of Variance Loss}

\begin{figure}[t]
\centering
\includegraphics[width=\linewidth]{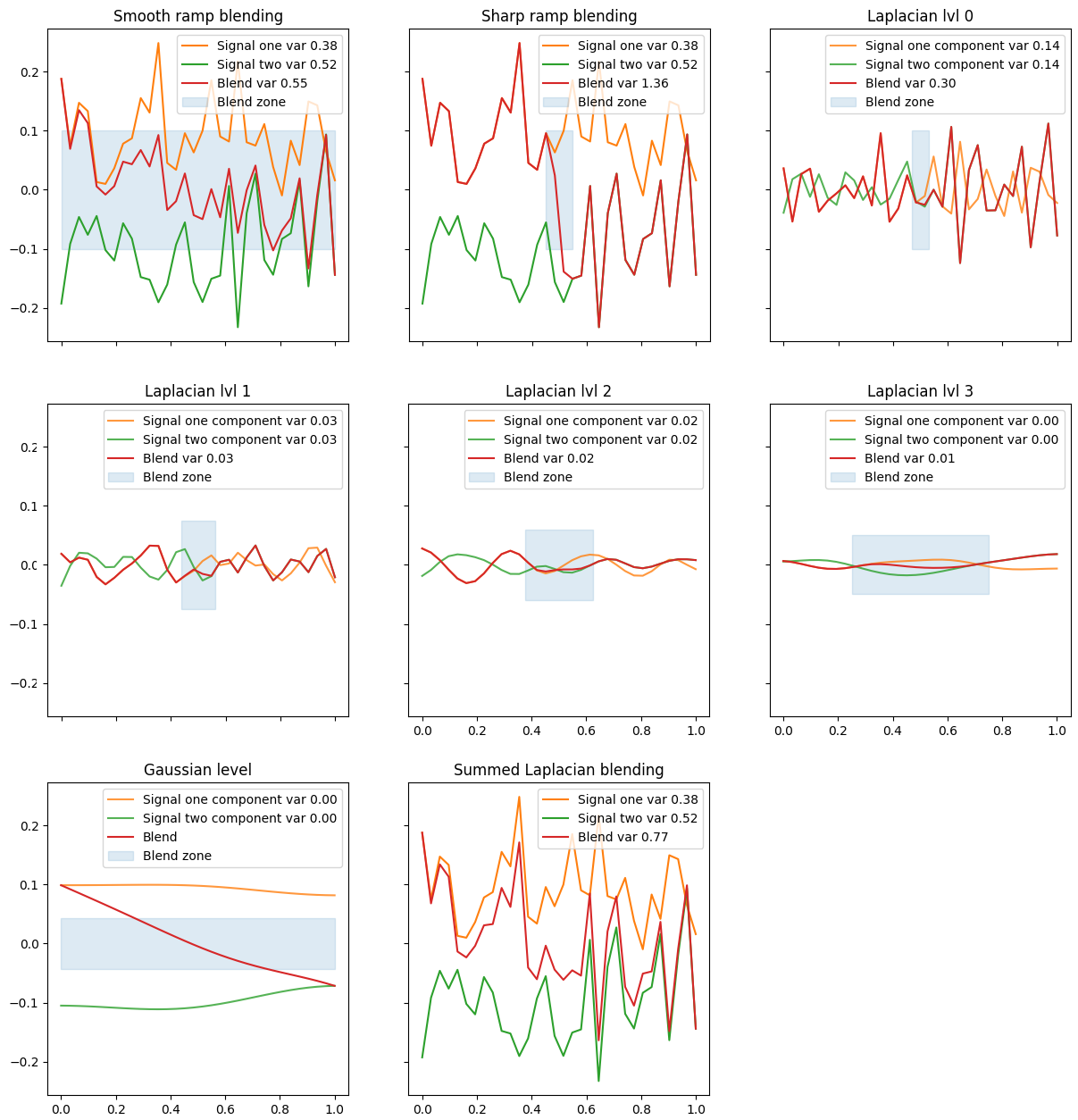}
\caption{A toy 1D example demonstrating how, with Laplacian blending, regions of variance loss are distributed between different levels.}
\label{diag_1d}
\end{figure}

If the Laplacian levels are uncorrelated, we can write
\begin{equation}
\mathit{Var}(X) \approx \mathit{Var}(L_{X0}) + \mathit{Var}(L_{X1}) + \mathit{Var}(L_{X2}) + \cdots + \mathit{Var}(G_{Xn}).
\end{equation}
In the case of a perfect Fourier decomposition, this equality is strict from Parseval's theorem.

In practice, the correlation of the Laplacian levels depends on the quality of the used filters and the specific signals (for example, due to insufficient filtering and aliasing).
We analyze the impact of the used filters in Section~\ref{sec_filtering_difference}, but for now, we assume that the correlation is small:
\begin{align}
\mathit{Cov}(L_{Xk}, L_{Xm}) &\approx 0,\\
\mathit{Cov}(L_{Xk}, G_{Xn}) &\approx 0.
\end{align}

Blending Laplacian pyramid levels with different radii, we distribute the variance reduction between different Laplacian levels and over different area sizes.
We show a toy 1D linear blending example in Figure~\ref{diag_1d}.
Two different 1D signals are blended, and a sharp linear transition region causes a visible discontinuous ``jump,'' while a smooth linear transition reduces sharpness and the original signal features.
Conversely, Laplacian blending distributes blending across different frequency content levels and region sizes.
For instance, the highest signal frequencies get attenuated only over very small regions.

The exact variance and contrast loss depends on the spectral content of the blended images and is always between the sharpest (single texel) and the widest blending radii.
This causes a smaller contrast or high-frequency detail loss than a wide blending radius while preserving its natural perceptual smoothness and lack of visible discontinuities.

\enlargethispage{6pt}

\section{Controlling the Behavior}
The proposed method does not require per-texture parameter tuning for robust behavior, but a few parameters and the filtering kernel choice impact its visual appearance.
\subsection{Laplacian Pyramid Level Count}
\begin{figure}[b]
\centering
\includegraphics[width=\linewidth]{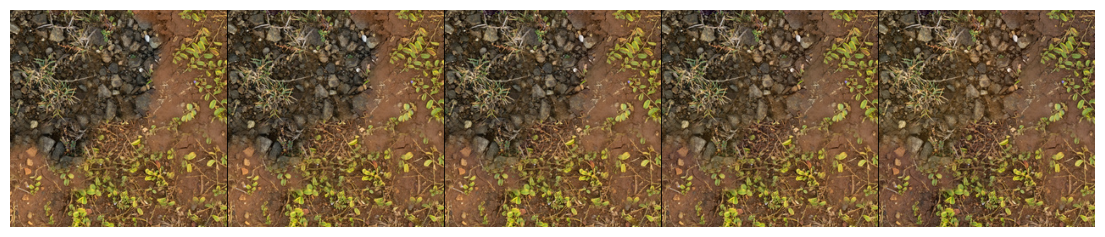}
\caption{From left to right: Blending with three to seven Laplacian levels.
A higher level count increases the smoothness and perceived continuity of the transition, but all preserve sharp contrast and high-frequency features without visible ghosting. The transitions above five Laplacian levels are very smooth, losing some of the distinct identity of the source textures. This can be both an advantage or undesirable from an artistic perspective depending on the use case.}
\label{levels_3_to_7}
\end{figure}

The Laplacian pyramid level count is the most important parameter we propose to expose for artistic appearance control.
It defines the effective radius of the transition. We suggest three to four levels as a practical default value.
Above five levels, the high transition radius causes average colors of different textures to blend and some of their unique appearance identity to be lost (Figure~\ref{levels_3_to_7}).
We note, however, that even such wide radius blending still doesn't show visible ghosting, and while the color contrast gets lower, the local contrast and high-frequency features are intact.

While we propose to use by default the same level of the mask Gaussian pyramid as the levels of the blended Laplacian pyramid, $0,\ldots, n$, it is possible to use biased further levels $k,\ldots, (n+k)$ for more aggressive blending with fewer levels.
However, this can lead to minor ghosting, as it becomes similar to direct linear blending with a larger radius.
\subsection{Prefiltering and Upsampling Kernel}\label{sec_filtering_difference}
There are many ways to construct a Laplacian pyramid.
For example, it can be either decimated (where each next level is a lower resolution and often constructed through subtracting progressively more low-pass-filtered and decimated images) or undecimated using bandpass filters.
We focus on the decimated case, as it allows for an efficient, practical, and low-memory-use implementation.
Decimated Laplacian pyramid behavior and each level's contents depend on both the filter used before decimation and the level upsampling filter:
\begin{equation}
L_{xk} = G_{xk}-F_{\mathit{up}}\left((F_{\mathit{down}}(G_{xk})\downarrow)\uparrow\right),
\label{laplac}
\end{equation}
where $F_{\mathit{down}}$ is a downsampling filter, $F_{\mathit{up}}$ is an upsampling filter, $\downarrow$ symbolizes a decimation operation (decreasing the resolution by dropping every other pixel in each dimension), and $\uparrow$ symbolizes the resolution increase operation by a factor of two by zero-insertion in each dimension prior to application of an upsampling $F_{\mathit{up}}$ filter.
Later in this text we will use the notation $\uparrow_{k}$ for a resolution increase by a $2^k$ factor.

Recommending a good and efficient filter is beyond the scope of this work, but we will analyze two commonly used downsampling filters in computer graphics.
The first is a box filter, often used for mipmap chain construction due to implementation simplicity, similar behavior to trilinear filtering, and the lowest possible cost---a single texture tap when using bilinear hardware samples.
The second one is the Lanczos2 filter~\cite{duchon1979lanczos}, a sinc windowed sinc filter with a very sharp frequency response given a relatively small spatial support ($4\times 4$ in the case of Lanczos2 2D downsampling).
The Lanczos2 filter demonstrates good low-pass filtering but tends to produce perceptual sharpening and some ringing.
Similarly to the downsampling filter, the upsampling filter choice can impact the visual results, but we focus on the most common bilinear upsampling filter due to its very small computational cost and hardware filtering support on the GPU.
\begin{figure}[t]
\centering
\subfloat[Box-filter Laplacian Pyramid]{\includegraphics[width=0.8\linewidth]{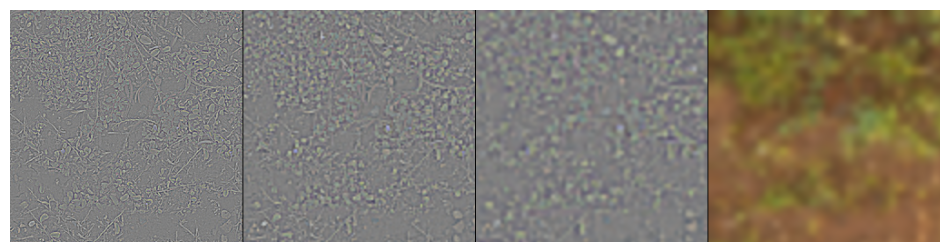}}\\
\subfloat[Lanczos4-filter Laplacian Pyramid]{\includegraphics[width=0.8\linewidth]{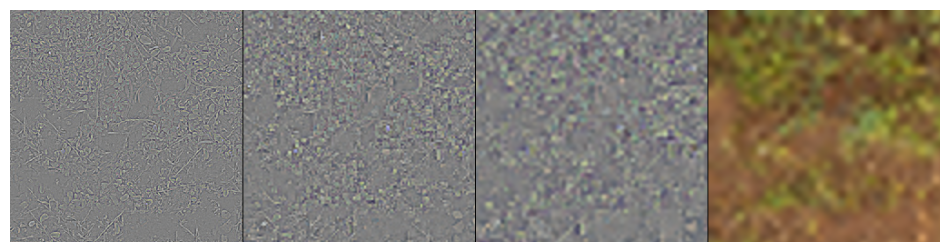}}
\caption{The impact of used downsampling filters on Laplacian pyramid contents. The Lanczos filter preserves more high-frequency contents in lower levels, including the final Gaussian level.}
\label{laplacian_filters}
\end{figure}

\begin{figure}[t]
\centering
\subfloat[Box-filter Laplacian Pyramid blending]{%
  \makebox[0.5\linewidth][c]{\includegraphics[width=0.4\linewidth]{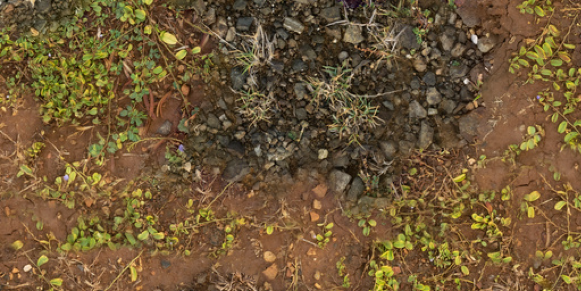}}%
}
\subfloat[Lanczos4-filter Laplacian Pyramid blending]{%
  \makebox[0.5\linewidth][c]{\includegraphics[width=0.4\linewidth]{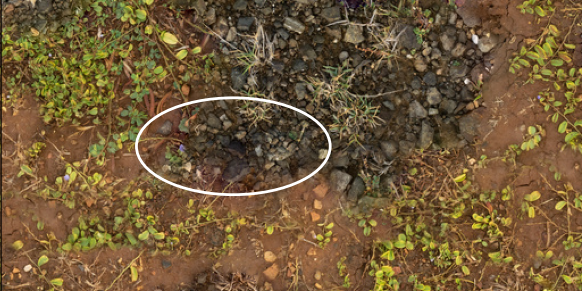}}%
}
\caption{Different downsampling filters and Laplacian pyramid creation methods produce a different visual outcome.
Lanczos2 results are sharper but produce over-darkening and overshoots in some regions: an example in the right image is marked with a yellow oval.}
\label{filtering_difference}
\end{figure}

Those two downsampling filters produce different Laplacian decompositions (Figure~\ref{laplacian_filters}).
The Lanczos filter separates filtered levels better, and thus, each Laplacian difference has more energy.
The filter also improves the sharpness of the final Gaussian level but creates some over- and undershoots.

This difference contributes to different visual outcomes of Laplacian pyramid blending (Figure~\ref{filtering_difference}).
Lanczos2 Laplacian pyramid blending is sharper but produces over-darkening and overshoots in some regions.
While noting the difference and that one filter might be subjectively preferred over the other, we conclude that the proposed method works well with either method.

\section{Practical GPU Implementation}
Creating, blending, and filtering multiple image pyramids might seem costly, but we propose an efficient, GPU-friendly implementation through two simple modifications of the core algorithm.
Instead of constructing a Gaussian pyramid, we use an existing texture mipmap chain.
Instead of explicitly constructing a Laplacian pyramid, we propose an approximation using difference of Gaussians.
We construct the full-resolution Laplacian level in place from two existing low-resolution texture mip levels:
\begin{equation}
L_{xk} \approx F_{\mathit{up}}\left((G_{xk})\uparrow_{k}\right)-F_{\mathit{up}}\left((G_{xk+1})\uparrow_{k+1}\right).
\label{approx_laplac}
\end{equation}
This approximation allows for the in-place construction of Laplacian levels in the final shader when a blended texture is read.
Listing~\ref{code} presents an example application with four levels.
With the proposed implementation, there is no memory storage overhead or precomputation--- assuming that textures already have mipmaps. This code scales to regular, pointwise linear blends---when \lstinline{NUM_LEVELS} is zero.

To sample $n$ Laplacian levels and an additional Gaussian level, we take just $n+1$ samples---where all the additional samples come from lower mipmaps (with a negligible bandwidth/cache cost).
Other than mipmap sampling, the arithmetic cost of our method is just multiply-adds and multiplies: the same as regular blending, but multiplying it $n+1$ times and accumulating all of the blended levels.

The whole cost of the method is $n+1$ times more samples for a given desired level and $n$ blends and adds.
In practice, the cost of additional samples doesn't need to scale linearly and depends on the hardware architecture.
The final GPU performance cost depends on various factors: the utilization of the texture unit, memory bandwidth, cache sizes, register usage, and arithmetic operations.
Even when computing the Laplacian pyramid from all the mip levels, the maximum used memory bandwidth is $133\%$ of the original cost (total mipchain pixel count).

The whole method is presented in Listing~\ref{code}. It is worth noting that in GLSL the third parameter of  the \lstinline{texture2D} method is the level-of-detail bias, which we will use in Section~\ref{mipmapping}.

\begin{lstlisting}[language=C, float=t, caption=Example implementation.,label=code,morekeywords={vec4}]
#define NUM_LEVELS 4
vec4 tex0_levels[NUM_LEVELS+1];
vec4 tex1_levels[NUM_LEVELS+1];
vec4 mask_levels[NUM_LEVELS+1];

for (int i = 0; i < NUM_LEVELS+1; i += 1) {
    tex0_levels[i] = texture2D(tex0, uv, float(i));
    tex1_levels[i] = texture2D(tex1, uv, float(i));
    mask_levels[i] = texture2D(mask, uv, float(i));
}

vec4 blended = vec4(0.0);
for (int i = 0; i < NUM_LEVELS; i += 1) {
    vec4 tex0_laplace = tex0_levels[i] - tex0_levels[i+1];
    vec4 tex1_laplace = tex1_levels[i] - tex1_levels[i+1];
    blended +=  tex0_laplace * (1.0 - mask_levels[i]) +
                tex1_laplace * mask_levels[i];
}
// Gaussian level.
vec4 tex0_gauss = tex0_levels[NUM_LEVELS];
vec4 tex1_gauss = tex1_levels[NUM_LEVELS];
blended +=  tex0_gauss * (1.0 - mask_levels[NUM_LEVELS]) + 
            tex1_gauss * mask_levels[NUM_LEVELS];
\end{lstlisting}

We report example timings in Table~\ref{performance}, measured by rendering in $3840 \times 2160$ pixels resolution, blending two textures on an NVIDIA RTX 4090 GPU, and repeating the blending 1000 times with changing UVs and averaging the overhead.
The typical Laplacian level count that produces smooth but sharp blends is three to five, for which the proposed method has a minimal runtime performance impact.

\begin{table}[t]
\centering
\begin{tabular}{@{}llllll@{}}
\toprule
Levels        & 1--3            & 4     & 5     & 6     & 7     \\ \midrule
Overhead (ms) & Not observed & 0.087 & 0.113 & 0.125 & 0.134 \\ \bottomrule
\end{tabular}
\caption{Performance overhead of the presented method in 4K resolution.}
\label{performance}
\end{table}

\subsection{Minification and Mipmapping}
\label{mipmapping}
The description of our method so far assumes that texture blending happens at the full resolution of the textures (the finest mip level).
This is sufficient for applications such as caching blending using virtual texturing, but would pose a problem on perspective-projected 3D assets requiring varying minification levels.

We can address this limitation with a simple modification of our algorithm.
We begin by observing that the spectral contents of a mip level $k$ are low-pass filtered image frequencies from the original full-resolution texture higher than its Nyquist levels.
From Equations~\eqref{laplac} and~\eqref{approx_laplac} we see that this is also equivalent to zeroing out the Laplacian levels $0,\ldots, k-1$ while keeping the further and coarser Laplacian/Gaussian pyramid levels.

This translates to two straightforward modifications of the code in Listing~\ref{code}. 
First, the desired mip level $k$ has to be queried using the \lstinline{textureQueryLod} method and the fetched levels start at $k$ instead of zero.
Second, the number of blended Laplacian levels and the index of the selected Gaussian level (lines 12 and 20) are set to be equal to \lstinline{max(NUM_LEVELS - k, 0)}.
This is functionally equivalent to the code in Listing~\ref{code} when no minification is present, drops the Laplacian level blending when the texture is minified beyond the coarsest blending level, and partially blends Laplacians in between.

However, we note that the results of alpha-blending minified textures are not the same as those of minifying alpha-blended textures, irrespective of the use of our method.
This is caused by the nonlinear nature of alpha mask multiplication and shading applied to material textures described by \citet{pharr2024filtering}, and our method is compatible with their family of stochastic filtering techniques.

\subsection{Optimization: Level Skipping}
\begin{figure}[t]
\centering
\includegraphics[width=\linewidth]{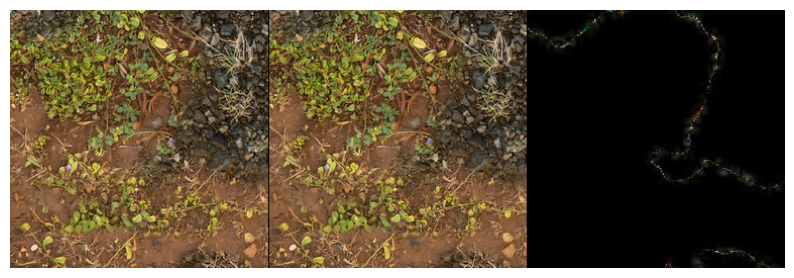}
\caption{Left: Original algorithm, four levels. Middle: Level skipping (two levels computed). Right: Absolute difference amplified $5\times$.}
\label{level_skipping}
\end{figure}

If the cost of the proposed method is too high (for instance, when blending multiple textures per material, or on mobile devices), one can use a further approximation for Laplacian creation by skipping the mip levels: for example,
\begin{equation}
\widehat{L}_{xk} \approx G_{xk}-F_{\mathit{up}}((G_{xk+2})\uparrow_{k+2}).
\end{equation}
Using such a modified definition, only Laplacians and Gaussian $\widehat{L}_{x0}, \widehat{L}_{x2}, \widehat{L}_{x4}, \ldots$ need to be computed and blended.
This reduces the cost overhead to $\frac{N}{2}+1 $ more samples and evaluations.
It's worth noting, however, that this changes the visual appearance and leads to some minor quality loss (less preservation of sharp features and more ghosting) presented in Figure~\ref{level_skipping}.

\subsection{Dynamic Blend Mask Levels}
\begin{figure}[b]
\centering
\subfloat[\label{dyn_linear}Linear mask]{\includegraphics[width=0.19\linewidth]{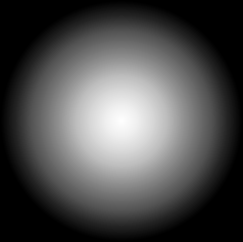}}\hfill
\subfloat[Approx. blend\\mask level 0]{\includegraphics[width=0.19\linewidth]{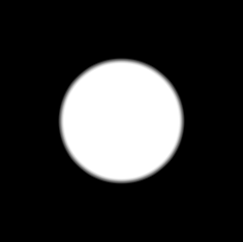}}\hfill
\subfloat[Approx. blend\\mask level 1]{\includegraphics[width=0.19\linewidth]{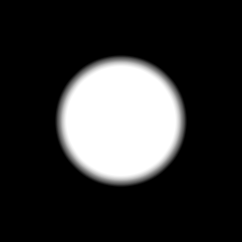}}\hfill
\subfloat[Approx. blend\\mask level 2]{\includegraphics[width=0.19\linewidth]{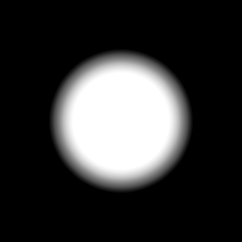}}\hfill
\subfloat[Approx. blend\\mask level 3]{\includegraphics[width=0.19\linewidth]{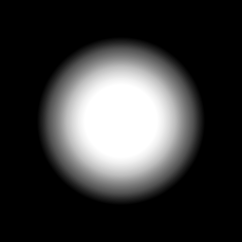}}
\caption{An example of creating different approximate Gaussian levels of blend masks dynamically from a source linear texture \protect\subref{dyn_linear}.}
\label{dynamic_masks}
\end{figure}

We propose a second modification to our technique that eliminates the need to create a Gaussian pyramid of the mask texture.
This modification can efficiently create dynamic and changing masks or remove the need to sample multiple mask texture mip levels.
In this technique variation, we use smooth masks resembling alpha maps or distance fields---they can be either procedural or stored in textures.
We obtain an approximation of the Gaussian level $n$ through clamped remapping, such as rescaled thresholding:
\begin{equation}
G_{mn} \approx \text{clamp}\left(\frac{G_{m0}-t}{s 2^n}+0.5,0,1\right),
\end{equation}
where $t$ is the threshold of the transition center and $s$ is a scale that depends on the values stored in the texture (for instance, texture resolution if the values are stored in the normalized $\left[0,1\right]$ range). See the example in Figure~\ref{dynamic_masks}.

\section{Applications}
The proposed method of Laplacian texture blending can be applied in different applications requiring smooth transitions between multiple textures in a real-time rendering context.
We discuss two main applications.

\subsection{Material Layering} The material authoring process often involves layering and blending multiple different textures of base materials~\cite{neubelt2013crafting,substanceblending}. The proposed method can be included in the material creation toolset as one of the blend modes. 

The material authoring process can either create dynamic, real-time materials or bake them into textures, and our method is compatible with both workflows.
The lack of precomputations makes it especially attractive for so-called \textit{uber materials}, where a single material can use different parameters and texture sets, sometimes changed in real time.
An example of dynamic adjustments can be a dynamic weather or season system in a rendering engine, when an animated mask progressively changes, revealing or covering a different material. 

\begin{figure}[b]
\centering
\includegraphics[width=\linewidth]{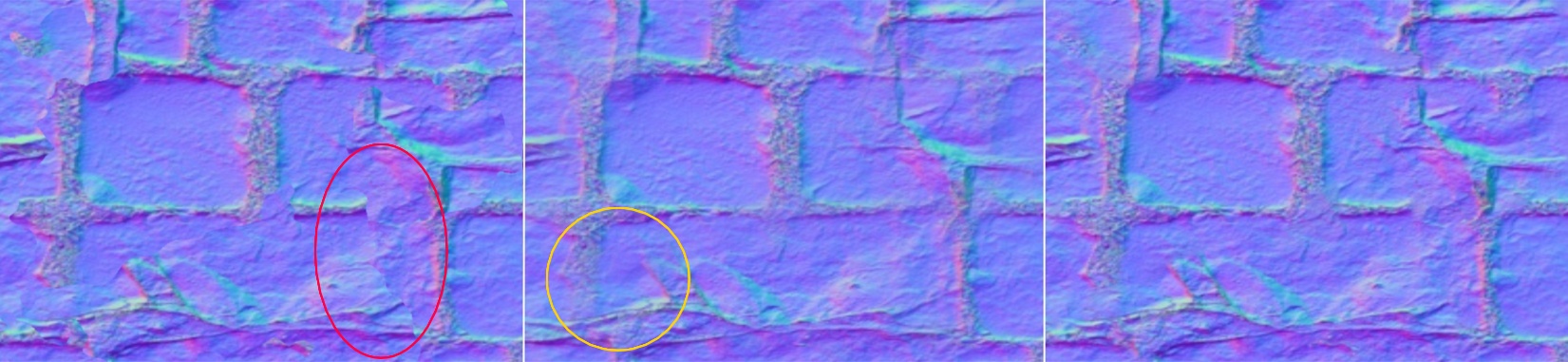}
\caption{Blending two unrelated normal maps. Left: A narrow linear blend results in visible seams (red oval). Middle: A wide linear blend attenuates and blurs the blended normal map details toward the normal pointing up (yellow circle). Right: A Laplacian pyramid blend preserves normal map details while not producing surface discontinuities.}
\label{normals_blend}
\end{figure}

Typically, materials comprise multiple textures of different BRDF properties and not just color information.
Our method aims to preserve gradients present in the image.
It makes no other assumptions about the color, its distribution, perceptual space, or the map semantics and thus works on any other type of property map.
For example, harsh transitions of blended normal maps can produce visible surface and lighting discontinuities.
We show an example of normal map blending in Figure~\ref{normals_blend}, where our method demonstrates the same perceptual smoothness of wide blends while preserving the fine details and edges.

\begin{figure}[t]
\centering
\includegraphics[width=\linewidth]{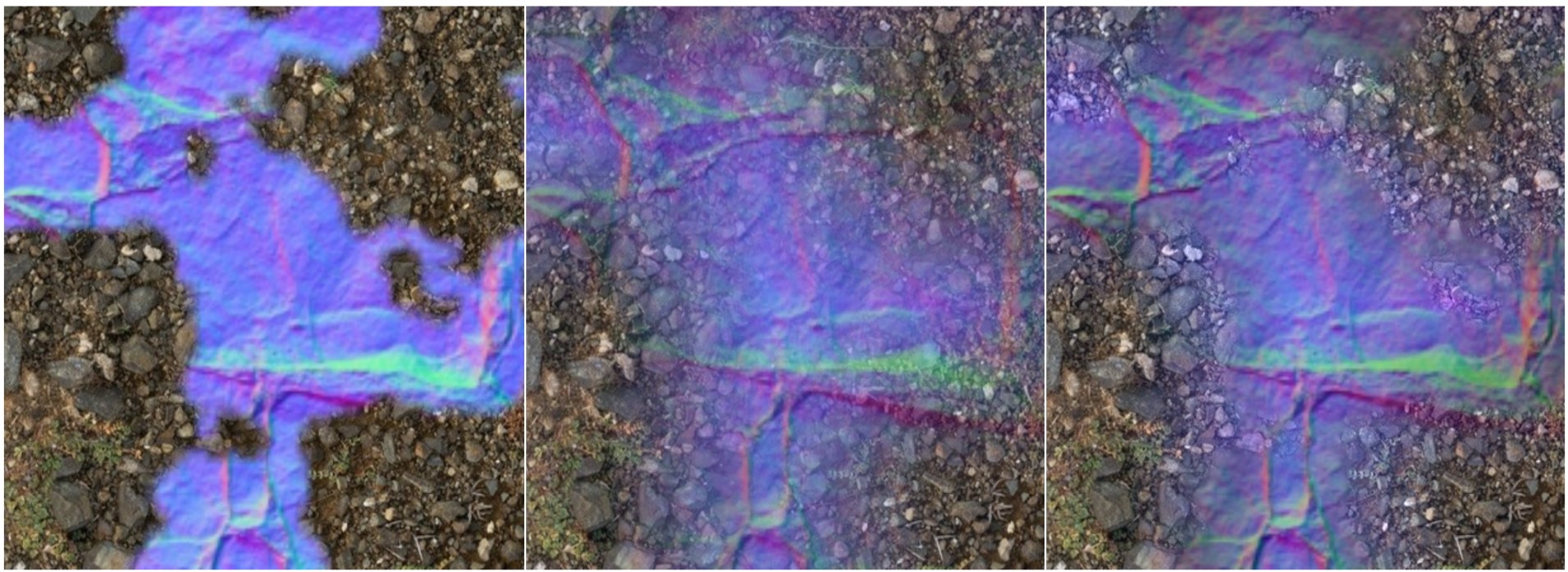}
\caption{Extreme blending of unrelated content. Left: Narrow linear blend. Middle: Wide linear blend. Right: Laplacian pyramid blend, which creates color blends as wide as the wide linear blend while preserving contrast and avoiding feature overlap and ghosting.}
\label{extreme_blend}
\end{figure}

As an additional experiment to showcase the strength of our method in this scenario, we demonstrate blending two completely unrelated textures in Figure~\ref{extreme_blend}.
Our method blends colors similarly to wide blends while preserving local features without visible ghosting.

\subsection{Procedural Texture Generation and Texture Tiling} 
\begin{figure}[b]
\centering
\includegraphics[width=\linewidth]{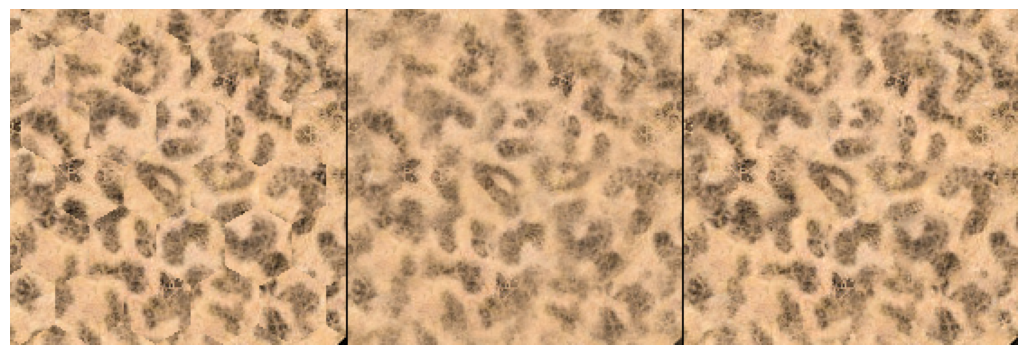}
\caption{Left: Hard tiling without transition reveals an obvious tile structure. Middle: Soft blending causes ghosting and contrast loss. Right: Laplacian blending retains most contrast and appearance while not producing visible tiles.}
\label{hex_tiling_example}
\end{figure}

Texture synthesis and hexagonal tiling literature~\cite{burley2019histogram,heitz2018high,Mikkelsen2022Hex} inspired our work, and our method is designed to work in such a scenario.
With hex tiling, every evaluated pixel blends three textures based on their distance from hexagon edges.
Our method is compatible with such a setup and, similarly to the original hexagonal tiling work, doesn't require sampling hexagonal masks, as the blending weights can be determined analytically for every level.
We present an example in Figure~\ref{hex_tiling_example}.
The advantages of our method are high-quality results, no need for precomputations, and no need for empirical tuning.
The biggest disadvantage is the increased cost. The base hex tiling method requires three samples for each material texture in the blended region;
our method increases it by a level-count-dependent factor.

\section{Limitations}

\vspace*{-3pt}
\subsection{Performance Cost}
The main limitation of the proposed method is the increased runtime cost: additional samples from lower mipmaps and arithmetic operations.
While additional samples use lower mipmaps and are more likely to be localized in the L1/L2 cache without using more memory bandwidth, it is possible to saturate the texture unit with too many requests.
We recommend profiling and evaluating the proposed method's cost, especially when blending multi-channel physically based rendering materials and more than two materials per pixel.
For particularly complex but common scenarios like terrain material blending, an alternative can be using the proposed method to blend into a cached virtual texture~\cite{chen2015adaptive}.

\vspace*{-3pt}

\subsection{Possibility of Overshooting and Haloing} 
Blending Laplacians independently and at different rates can produce halos or overshooting (Gibbs phenomenon) and negative values.
This was reported as a problem in HDR exposure fusion literature with proposed heuristics and workarounds~\cite{hasinoff2016burst}.
We have not observed those problems using a bilinear upsampling filter in the evaluated examples.
Furthermore, using only a few Laplacian levels limits the maximum potential halo size.
However, we recommend clamping the blended textures to the original texture value range $[0, 1]$.

\vspace*{-3pt}

\subsection{Possibility of Visible Aliasing on Animated Content}
If the material blending mask or blended textures are animated, imperfect filtering during the Laplacian construction can lead to aliasing.
While high-quality filters would minimize this problem, it can occur when using a typical box filter and bilinear upsampling (Section~\ref{sec_filtering_difference}).
We note that it is similar to any dynamic mipmap creation, like for the bloom effect, and real-time rendering literature proposes using stronger low-pass filters while keeping the low-cost upsampling filters~\cite{jimenez2014next}.

\vspace*{-3pt}

\section{Conclusion}
In this work, we proposed an efficient, GPU-friendly adaptation of Laplacian image blending for real-time rendering applications---semi-procedural materials, material texture layering, and example-based synthesis like hex tiling.
The proposed method can be used both in a fully automated setting and presented to the artists as an additional tool for creating and blending textures and materials.

Perceptual characteristics of Laplacian pyramids have been used in image processing and computational photography literature for many years.
Meanwhile, real-time graphics use image pyramids almost exclusively for performance acceleration and processing some effects at lower resolutions.

While different performance and memory storage requirements between different computer science domains require re-designing and approximating key components of any adapted algorithm, we believe that computer graphics can benefit from further adoption of computational photography techniques such as multi-level signal decomposition and nonlinear blending and filtering.

\small
\bibliographystyle{jcgt}
\bibliography{paper}

\section*{Index of Supplemental Materials}
A WebGL demo is provided as supplementary material.

\noindent Download: \vspace*{-1ex}
\begin{itemize}
  \item \url{https://jcgt.org/published/0014/01/02/supplement_demo.zip}
\end{itemize}
\noindent Run live: \vspace*{-1ex}
\begin{itemize}
  \item \url{https://jcgt.org/published/0014/01/02/supplement_demo}
\end{itemize}

\section*{Author Contact Information}

\hspace{-2mm}\begin{tabular}{p{0.5\textwidth}p{0.5\textwidth}}
Barlomiej Wronski \newline
NVIDIA, USA \newline
\href{mailto:bwronski@nvidia.com}{bwronski@nvidia.com}

\end{tabular}

\afterdoc

\end{document}